\documentclass[
]{ceurart}

\usepackage{wrapfig} %
\usepackage{caption} %

\usepackage{listings}
\begin{document}

\copyrightyear{2025}
\copyrightclause{Copyright for this paper by its authors.
  Use permitted under Creative Commons License Attribution 4.0
  International (CC BY 4.0).}

\conference{INRA 2025: 13th International Workshop on News Recommendation and Analytics in Conjunction with ACM RecSys 2025,
  September 22--26, 2025, Prague, Czech Republic}

\title{Towards Multi-Aspect Diversification of News Recommendations Using Neuro-Symbolic AI for Individual and Societal Benefit}

\author[1]{Markus Reiter-Haas}[%
orcid=0000-0001-9852-8206,
email=reiter-haas@tugraz.at,
url=https://iseratho.github.io/,
]
\cormark[1]
\address[1]{
Graz University of Technology, Institute of Human-Centred Computing, 
Sandgasse 36/III, 
8010 Graz,
Austria}

\author[1]{Elisabeth Lex}[%
orcid=0000-0001-5293-2967,
email=elisabeth.lex@tugraz.at,
url=https://elisabethlex.info/,
]

\cortext[1]{Corresponding author.}

\begin{abstract}
News recommendations are complex, with diversity playing a vital role. So far, existing literature predominantly focuses on specific aspects of news diversity, such as viewpoints. In this paper, we introduce multi-aspect diversification in four distinct recommendation modes and outline the nuanced challenges in diversifying lists, sequences, summaries, and interactions. Our proposed research direction combines symbolic and subsymbolic artificial intelligence, leveraging both knowledge graphs and rule learning. We plan to evaluate our models using user studies to not only capture behavior but also their perceived experience. Our vision to balance news consumption points to other positive effects for users (e.g., increased serendipity) and society (e.g., decreased polarization). 
\end{abstract}


\maketitle

\section{Introduction}

News recommender systems (NRS) have become increasingly popular and important for news consumption~\cite{karimi2018news}. An important concern is the news diversity in recommendations~\cite{bernstein2020diversity}, as diverse news consumption corresponds to a more balanced information diet~\cite{kulshrestha2015characterizing}. However, compared to other recommendation areas, the news domain has unique challenges, especially temporal considerations due to articles being short-lived and typically consumed at most once per user, which leads to increased sparsity~\cite{raza2022news}. 
Besides, NRS also need to deal with limited information about the users, evolving categorical preferences, a broad range of topics being covered, content being framed in certain ways, ideological blind spots in consumption, media biases of the news sources, as well as ethical and regulatory considerations. As news plays a vital role in opinion formation~\cite{kuhne2015emotional} and democracies~\cite{helberger2021democratic}, the diversification in NRS is especially relevant~\cite{vrijenhoek2021recommenders}. That is why works like \cite{draws2021assessing,mulder2021operationalizing} focus directly on diversifying viewpoints. In this paper, we argue that incorporating single aspects is not enough for news diversity and present multi-aspect diversity as an emerging research direction.

The conceptualizations of media and news diversity vary across literature comprising different subdimensions, such as topic and viewpoint diversity~\cite{loecherbach2020unified}. \citet{mattis2024nudging} introduce algorithmic nudges by re-ranking to increasing exposure diversity and presentation nudges via altering the user interface for deliberate changes in consumption diversity. \citet{bernstein2020diversity} outline that the role of news diversity differs from other recommendation domains, with the promotion of diverse content being vital in a democratic society. One particular observation from \cite{bernstein2020diversity} highlights that current models for NRS typically do not capture the multidimensionality of diversity, which is a driving motivator for the current paper.
Moreover, diversification in personalized NRS tends to focus on specific recommendation modes, typically targeting either the (sequential) history of clicked news or the list of recommended news~\cite{wu2023personalized}. For instance, \citet{wu2022end} trains an end-to-end diversity-aware news recommender based on semantic similarity in the recommendation list.
Therefore, \citet{wu2023personalized} emphasize diversity-aware NRS concerning both modes, as well as fine-grained diversification that goes beyond content, as a promising research direction. In this vein, \citet{kaya2018accurate} considers subprofiles, rather than item features, for intent-aware diversification. 

There have been important first strides towards multi-aspect news diversity. \citet{zhou2020diversifying} adapt the biology-inspired Simpson's diversity index for multi-aspect diversity in search results. \citet{li2025d} rely on random walks using normative distributions for transparent diversification of news candidates. \citet{vercoutere2024improving} uses concept graphs for extending the user profile improving diversity and the potential to increase user satisfaction according to their user study. The MANNeR model \cite{iana2024train} tackles multi-aspect diversity for NRS by incorporating a pairwise similarity matrix. In their formalization, diversity is defined as the normalized entropy over the aspects' distributions and models aspect via separate labels per aspect in recommendation list.

Our contributions in this idea paper are two-fold. 
First, we introduce the problem of \emph{multi-aspect news diversity in four distinct news recommendation modes}, where we also model similarities between labels of the same set of aspect. 
Thereby, our work expands the body of literature on diversity in recommendations~\cite{kunaver2017diversity} by thoroughly examining multi-aspect diversification of news, also in light of distinct novel modes like LLM-generated news summaries. 
Second, we aim to incorporate this expanded problem formalization to directly in the model architecture, as the training objective, and as transparent rules in deployed systems. Herein, we propose a research direction for \emph{diversifying based on both symbolic and subsymbolic AI using knowledge graphs and rule learning} together with user study evaluations to gauge their beneficial impact. While several neuro-symbolic approaches exist~\cite[refer to][]{carraro2025neuro}, to the best of our knowledge, none of them provide a universal solution to diversify arbitrary combinations of aspects with both individual and societal benefits in mind.

\section{Problem: Multi-Aspect News Diversity}
\label{sec:problem}

\begin{wrapfigure}{r}{.5\linewidth}
    \centering\includegraphics[width=\linewidth]{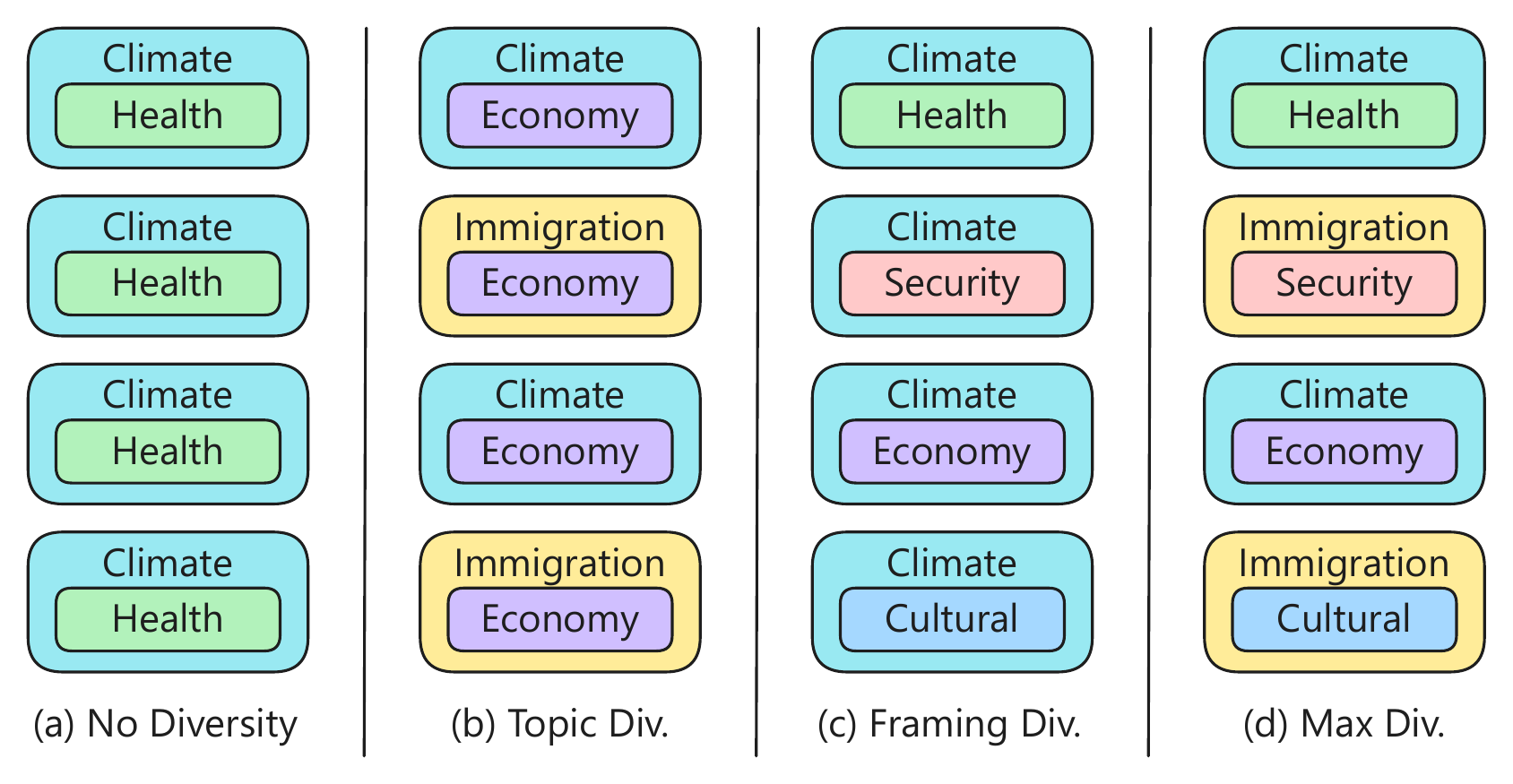}
    \captionsetup{format=plain}%
    \caption{Examples of Multi-Aspect Diversity in Lists. (a) A list of all similar items has the minimum diversity of $div=0$. (b) A diversified list regarding topic, but no framing diversity ($div=0.33$). (c) A diversified regarding framing, but no topical diversity ($div=0.42$). (d) A maximally diversified list regarding both topic and framing ($div=0.75$).
    }
    \label{fig:multilevel}
\end{wrapfigure}

Let $D=(d_1,d_2,\ldots,d_n)$ be an arbitrary collection of documents. Each document comprises data and metadata, while $dist: D \times D \rightarrow[0,1]$ is a distance function between pairs of documents $(d_i,d_j)$; we denote $dist(d_i,d_j)$ as their distance value. Furthermore, we assume the distance function satisfies $dist(d_i,d_i)=0$ and is symmetric: $dist(d_i,d_j)=dist(d_j,d_i)$. The diversity of the whole collection can be measured as average distance between all pairs of documents with $div(D) = \frac{1}{|D|\cdot (|D|-1)} \sum_{\forall d_i, d_j \in D, d_i \neq d_j} dist(d_i,d_j)$. For simplicity, we assume each document $d_i$ consists of only a topic label $t_i$ and frame label $f_i$ as aspects: $d_i=(t_i,f_i), t_i \in T, f_i \in F$, where $T$ and $F$ represent the set of topic and frame labels, respectively. Note that topic and frame labels can be seen as operating at different levels, which can be parameterized with $\alpha$ in the distance function to account for topical and framing distance separately: $dist(d_i,d_j)=\alpha \cdot dist_t(t(d_i),t(d_j)) + (1-\alpha) \cdot dist_f(f(d_i),f(d_j))$. The functions $t(\cdot )$ and $f(\cdot )$ retrieve the topic and frame of the document, respectively. 

For illustration, we define two topics $T=\{\text{Climate},\text{Immigration}\}$ and four frames $F=\{\text{Health},\text{Cultural},\text{Security},\text{Economy}\}$ and set $\alpha=0.5$. We assign the values: $dist_t(\text{Climate},\text{Immigration}) = dist_f(\text{Health},\text{Security}) = dist_f(\text{Health},\text{Economic}) = dist_f(\text{Cultural},\text{Security}) = dist_f(\text{Cultural},\text{Economic}) =1$ and $dist_f(\text{Health},\text{Cultural})=dist_f(\text{Security},\text{Economic})=0.5$. This effectively means the frames Health/Cultural and Security/Economic bear some resemblance (e.g., similarity in ideology), while both topics and other framing pairs are seen as maximally dissimilar to each other.

Figure~\ref{fig:multilevel} visualizes different scenarios for diversity of four documents in a list without considering their order. Note that neither for topic nor framing the maximum diversity of $1$ can be reached in this scenario. The number of topics is smaller than the number of items in the list ($|T|<|D|$). While the number of frames is the same as the length of the list, not enough frames are completely dissimilar to each other: $max(\{|F_{dis}|,F_{dis} \subseteq F, \text{where } \forall f_i,f_j \in F_{dis}, f_i \neq f_j, dist_f(f_i,f_j)=1\})<|D|$. Therefore, in this example, only a list of length two (i.e., the minimum number of completely distinct labels per aspect) can achieve the global maximum diversity of $1$, where diversity is essentially reduced to item coverage.

The example showcases that even in this almost minimal example, diversifying news is far from trivial. In reality, the complexity of the problem increases tremendously just by scaling the number of aspects (e.g., adding news categories like sport or politics), the number of labels per aspect (e.g., additional topics being covered), and the size of the recommended list. Besides, the distance function is typically not so precisely defined, as several other factors also play a role. For instance, the Health frame in Climate could be semantically different than the Health frame in Immigration, operating at different hierarchical levels. Similarly, users could perceive diversity differently depending on the order, where an interleaved list might be seen as more diverse compared to a neatly divided list (e.g, first half Climate, second half Immigration). Furthermore, considering latent variables, such as derived from the content itself, further exacerbates these problems. 

We would like to underscore that for brevity we have discussed diversity as if it were a measure that is completely independent of users, contexts, or other (e.g., ethical) considerations. Of course, this assumption is rather unrealistic, which is why the maximum diversity tends to differ profoundly from the optimal diversity. Diversity must also consider whether the recommendation is relevant for the user in a particular context and whether it adheres to established norms. A user who does not like sports will not benefit from general sport recommendations but might still monitor specific Olympic results, while spreading misinformation is detrimental for society even if considered relevant for a singular user. Therefore, maximizing the diversity cannot be the sole goal of a system. For the purpose of this paper, we nonetheless focus on increasing diversity without addressing these additional constraints.

So far, we have considered diversity in lists, which is a typical mode for recommendations. However, recommendation modes can also have more nuanced characteristics, like a spatial position. Depending on whether the order (or spatial position, for that matter) influences the distance function, the scenarios and resulting diversity differ or stay the same. The extent of these influences also plays a vital role. Recommendations presented as tiles in a grid (e.g., MSN homepage) or as categorical carousels (e.g., Inoreader dashboard) might not fundamentally differ for diversity purposes from a simple list. On the other hand, recommendation modes like endless sequences of news (e.g., on social media platforms like X) or LLM-generated summaries (e.g., on Perplexity AI discover feature) might require fundamentally different approaches. 

\section{Modes of News Diversification}
\label{sec:modes}

\begin{figure}[tbp]
    \centering
    \includegraphics[width=\linewidth]{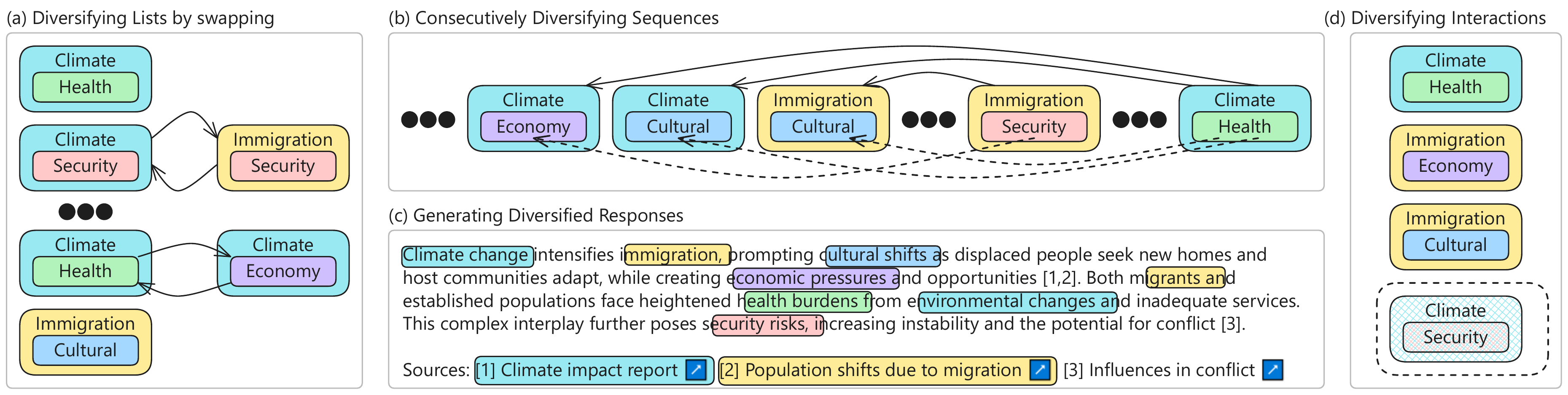}
    \caption{Diversifying news recommendations in different modes. (a) Lists can be diversified by swapping articles with others that have different aspects present. (b) Sequences need to consider items in a specific window, where the focus lies more on the similarity towards more recent items. Arrows denote similarity to previous items. (c) Summaries can be diversified at content or source-level. (d) Considering diversified interactions adds a new layer of complexity to the considerations, i.e., per interaction type such as likes or shares. The suggested interaction (shaded) maximizes both sharing and overall (including likes) diversity.}
    \label{fig:considerations}
\end{figure}

In this section, we discuss news diversification in four distinct recommendation modes 
: recommendation lists, sequences, summaries, and interactions (Figure~\ref{fig:considerations}). 
Note that these modes can be combined (e.g., lists of summaries or sequences of lists). For the present paper, we consider them strictly in isolation.

\subsection{Diversifying Lists}

Diversifying lists closely follows the problem description in Section~\ref{sec:problem}. Variants of this mode use UI elements like the previously mentioned grids, tiles, and carousels, but also cards or different submenus and subpages. A characteristic of list-like recommendations is that they can all be generated at the same time (e.g., at retrieval time or even offline) and then presented to the user at once. While the order (or location) is relevant for the user, for measuring diversity it is typically omitted and only evaluated globally on the whole list. Examples for this mode are individual news websites (often regional or specialized ones with limited content), digital versions of print media, or newsletters. A simple way to increase the diversity is by swapping items in the list with dissimilar ones as a post-processing step (Figure~\ref{fig:considerations}a). That is, find an item with an overrepresented trait (e.g., label) and replace it with a similar item that has an underrepresented trait in the same aspect instead, which can be repeated. Note that this might not lead to optimal lists and that suitable substitutions might not be available. Furthermore, the presented definition completely dismisses temporal considerations like the presence of a user history. Especially in the news domain, users want to consume novel items rather than old (or repackaged) articles that they have seen previously (i.e., items from their history).

\subsection{Diversifying Sequences}

Unlike lists, diversifying sequences necessitates temporal considerations. In sequences, recommendations are generated at multiple points and previous items need to be considered. Sequences can generate individual recommendations or sets of recommendations at once. Noteworthy variants are paginated lists or endless scrolling pages. Besides, sequences can often be grouped into distinct sessions, either explicitly (e.g., login and logout) or implicitly (e.g., based on gaps in the consumption). Examples for this mode are large-scale news aggregators, mainstream news websites, RSS feeds/blogs, and social media platforms. When considering the diversity, we can consider all items in a particular window of time, i.e., from a cut-off point in the past until the present, and treat this as a list. When generating new recommendations, consider the history of previous items in the window and the potential candidates, and recommend the ones that maximize the diversity (Figure~\ref{fig:considerations}b). Here, the temporal order plays a role. Consider the scenario where two items are equally valid due to a balance in the window. Then recommending the item is more dissimilar to recent items is preferable, as this improves the diversity once the window shifts. 

\subsection{Diversifying Summaries}

When generating summaries, only a single item (i.e., the summary) is recommended. Therefore, the properties of the item itself must be diversified (Figure~\ref{fig:considerations}c). This can happen at the source level or content level (or both). At the source level, the articles used to generate the summary should be diverse (similar to list diversity). At the content level, pieces of the content (such as keywords) should have diverse aspects. As summaries with nowadays large language models (LLMs) are generated sequentially, similar principles to diversifying sequences can be applied. Typically, diversity at the source level should lead to diversity at the content level as well. Therefore, a focus lies on diversifying the retrieval part in retrieval-augmented generation (RAG) to enhance the LLM's response. Examples for this mode are AI overviews (although typically not displayed for the most novel news), AI search engines, and AI support tools to summarize news pages. A variant of this type is a conversational agent that generates multiple responses. Again, the same principles as in diversifying sequences apply.

\subsection{Diversifying Interactions}

The three previous modes focus on presenting diversity to the user rather than trying to diversify their actual behavior. Typically, it is assumed that recommender systems suggest relevant items, which the user then consumes (e.g., clicks and then potentially reads). However, different interactions convey different semantics. For instance, a user might like one kind of content but share a different one (Figure~\ref{fig:considerations}d). Therefore, another user might observe (and incorrectly assume) a low diversity (e.g., biased) behavior, while a third user might have the completely opposite observation. For that reason, we need to consider the interactions in addition to the interacted items. This can be modeled with a weighted average across different interaction types. Diversified recommendations should consider the type of interaction in the globally interacted items.

\section{Proposed Research Direction}
\label{sec:direction}

Although the four listed recommendation modes require tailored solutions, we believe that at the core a common research direction for all of them emerges. 

\subsection{Algorithmic Advancements}

\begin{figure}[tbp]
    \centering
    \includegraphics[width=.9\linewidth]{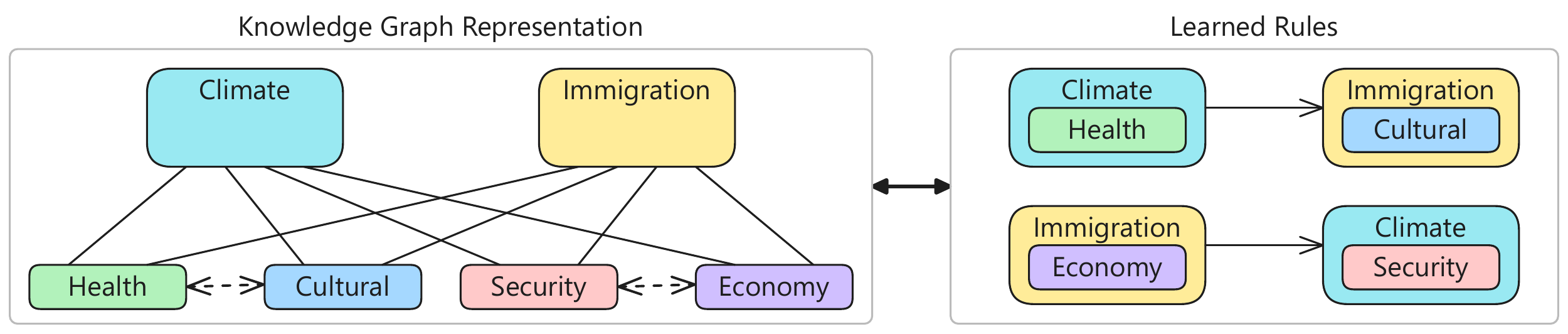}
    \caption{Integrating symbolic knowledge into news recommendations. Different aspects can be modeled as a knowledge graph (left) on which rules can operate (right).}
    \label{fig:graph}
\end{figure}

\leavevmode\indent
\emph{Novel Diversity Metrics. }
Diversity is often just assessed after the algorithm has been trained. Even so, established diversity metrics consider singular notions of similarity (e.g., regarding topics~\cite{ziegler2005improving} or viewpoints~\cite{draws2021assessing}), which do not explicitly capture nuanced differences such as in the framing of text. 
Instead, we call for generalized metrics that consider a set of relevant aspects as showcased in Section~\ref{sec:problem} to enhance existing diversity considerations. These aspects can be modeled as a knowledge graph to consider the interplay of various labels. In the guiding example, the documents can be represented by (hierarchical) knowledge graphs (depicted left in Figure~\ref{fig:graph}), which can be extracted with a topic and framing classifier from the content. Besides considering diversity separately, it can be incorporated with accuracy for a single target metric.

\emph{Knowledge-Graph-Enhanced Models. }
A naive methodological approach is to optimize for these metrics~\cite{li2024evaluation,wu2024result}. Furthermore, we can incorporate knowledge graphs into the news production~\cite[see ][for general and news-specific approaches, respectively]{opdahl2022semantic,guo2020survey}. Specific examples of knowledge-graph enhance models comprise considering similarity~\cite{puthiya2016coverage}, preference networks~\cite{xie2021improving}, concept graphs~\cite{vercoutere2024improving}, and including collaborative edges~\cite{liu2019news}. 
These knowledge graphs are then (jointly with the other content) fed into the system to (re-)rank the recommendations. However, instead of using them solely as auxiliary information in the input, we argue to exploit it as a regularizer in the loss (i.e., output of the model). Based on this, we can learn an inherent metric that captures the multi-aspect diversity. After bootstrapping it with some initial knowledge graphs, the aspects can automatically be extracted (i.e., derived) from the content in a self-supervised manner. Therefore, our approach combines subsymbolic representation learning with symbolic representations. Concerning the different modes, the application for lists is straightforward and sequential recommendations can use novel architectures like \mbox{xLSTM}~\cite{beck2024xlstm}. While sequential in nature, summaries would also benefit from retrieval-augmented generation for source diversity. Diversified interactions are more sophisticated and subject to data-driven reasoning like learned rules~\cite[e.g.,][]{manoharan2020intelligent}.

\emph{Incorporating Transparent Rules. }
Finally, and crucially, we plan to incorporate transparent rules that operate on the knowledge graphs~\cite{ma2019jointly}, e.g., Figure~\ref{fig:graph} right depicting examples rules. 
The rules can be set either (1.) globally, (2.) context-specific, (3.) per recommendation request, or (4.) derived from the user behavior. 1.~Globally defined rules are hardcoded in the system, e.g., to avoid spreading potential misinformation. Global rules are important for recommender platforms to adhere to regulations and internal goals. 2.~Context-specific rules can be set to affect a subset of recommendations, e.g., when a local event is relevant for people within a region. Being able to incorporate context-specific rules allows for quick adaptation of the system. 3.~Some parameters are set when retrieving the recommendation, e.g., when a filtering operation is applied by the user. This enables interactivity and control for the user. 4.~Certain patterns might be predominant, and explicitly learning them could increase system performance. Learning such rules could also lead to increased transparency in the system, as they can be leveraged for explainability.

\subsection{Evaluation Design}

The algorithms will be evaluated on established news datasets~\cite{kruse2024eb,wu2020mind,gulla2017adressa}. Unlike accuracy, which can be estimated offline by maximizing using historical behavior data, determining an ideal level of diversity is more challenging. Maximizing diversity could lead to adverse effects (e.g., backfire when exposed to opposite views~\cite{bail2018exposure}), which makes determining a ideal level of diversity more challenging. Also, the accepted level of diversity is user-dependent~\cite{wu2018personalizing}. One offline evaluation approach is to target a specific diversity level, e.g., derived from the data. Yet, aiming for the expected (i.e, mean) diversity per user would not lead to an increase thereof. Consequently, employing the recommender systems in an online scenario is necessary to determine the actual impact of algorithms. 

Our research plan is to determine the impact of the algorithms with user studies, once the offline evaluation shows promising results. We have the following online experiments in mind: using dedicated research platforms like POPROX~\cite{burke2025conducting} to recommend diversified lists, employing bots on social media platforms like BlueSky that users can follow to receive personalized sequential recommendations, collaborating with platforms (e.g., the Social Media Accelerator~\cite{bail2023we}\footnote{\url{https://www.polarizationlab.com/social-media-accelerator}}) to employ conversational AI agents to discuss news, and building a dedicated dynamic news system for exploring interactivity with rules. For brevity, we only illustrate the design for POPROX\footnote{See \url{https://poprox.ai/experimenter} for a platform description}. We aim to generate a daily newsletter for a few weeks using a diversified list based on articles from the Associated Press, where we perform the knowledge graph-induced computations for the recommendations while POPROX handles the user recruitment and content delivery. Importantly, POPROX enables users to answer a weekly survey, where we plan to ask general questions about their platform experience (e.g., their enjoyment and engagement), fatigue regarding certain kinds of news (e.g., politics), news habits (e.g., regarding news avoidance). Besides, specific questions on the news items themselves will be asked like whether any surprising news was included (i.e., serendipity) or what feeling a certain news item left the user with (i.e., positivity). We believe that diversified news can favorably influence the subjective perception of users. This is of great societal relevance, as straining the user with repetitive negative news could lead to avoidance and biased consumption patterns that subsequently lead to opinion polarization. We will conduct a randomized controlled trial where we split the user base into groups, with one group being the control, i.e., the standard newsletter, while the rest get the diversified news treatment(s).

\subsection{Research Vision}

We believe that diversifying news consumption is great for both societal and individual benefit. First, news recommender systems can indirectly act as bridging algorithms for depolarization. By delivering news that is more moderate to two opposite groups assuming extreme viewpoints, it can lead to higher agreement between them. For instance, by slightly reframing the topic of immigration towards a cultural perspective rather than a security issue (or vice versa). Especially, since it has been shown that news consumer tend to repeatedly consume similarly framed news~\cite{reiter2024framing}. 
Moreover, an audience that has a broader interest diversity, e.g., into a broader spectrum of news, acts as a safeguard for robust democracies~\cite{bednar2021polarization}. 
These benefits become more apparent when considering that news consumers might only be exposed to derivatives of news, such as summaries or commentaries, rather than the original sources. 

Second, news diversification has benefits for individual users. Related to that is serendipity~\cite{kotkov2016survey}, which is seen as unexpected but useful discoveries. Hence, it requires a broader spectrum of recommendations to include items that the users were unaware of. Therefore, we expect an elevated diversity to not only also increase serendipity but likewise positively influence their perception of the system in general. Besides passive improvements in metrics, users can become more engaged with the system when they have additional options to influence (by enabling/disabling rules to tweak individual preferences) and understand the (more transparent) recommendations. Finally, users could strive to optimize their own behavior. For instance, news recommendations could provide a score on their behavior with rewards for positive actions (like constructively engaging with articles of opposing viewpoints).

\section{Conclusion}
\label{sec:conclusion}

The problem of considering multiple aspects in news diversity is complex, and existing approaches remain insufficient. We present a novel research direction incorporating symbolic representations to diversify news in four distinct recommendation modes. Particularly in light of the increasing utilization of LLMs for news summarization and active user engagement through a variety of news interactions, it is imperative to address the issue holistically. Our research vision points to a future where people can once again discuss important societal topics based on the same foundational facts while also making personalized news consumption more pleasant.

\begin{acknowledgments}
  This research was funded in whole or in part by the Austrian Science Fund (FWF) 10.55776/COE12. 
\end{acknowledgments}

\section*{Declaration on Generative AI}
  During the preparation of this work, the author(s) used Gemini, LanguageTool in order to: Grammar and spelling check, Improve writing style, Paraphrase and reword. After using this tool/service, the author(s) reviewed and edited the content as needed and take(s) full responsibility for the publication’s content.

\bibliography{refs}

\end{document}